\documentclass[12pt,journal,compsoc, onecolumn]{IEEEtran}
\usepackage[]{graphicx,adjustbox, color}

\ifCLASSOPTIONcompsoc
\else
\fi
\ifCLASSINFOpdf
\else
\fi
\begin{document}
%\graphicspath{images/}

\title{Aqua Computing: Coupling Computing and Communications}

\author{Chathura~Sarathchandra Magurawalage,
        Kun~Yang, and~Kezhi~Wang%  <-this % stops a space
\IEEEcompsocitemizethanks{\IEEEcompsocthanksitem  The authors are with the school of Computer Science and of Electronic Engineering, University of Essex, CO4 3SQ, United Kingdom.\protect\\


% Uncomment this
E-mails: \{csarata, kunyang, kezhi.wang\}@essex.ac.uk}% <-this % stops a space
}

\IEEEcompsoctitleabstractindextext{%

\begin{abstract}
%\boldmath

The authors introduce a new vision for providing computing services for connected devices. It is based on the key concept that future computing resources will be coupled with communication resources, for enhancing user experience of the connected users, and also for optimising resources in the providers' infrastructures. Such coupling is achieved by Joint/Cooperative resource allocation algorithms, by integrating computing and communication services and by integrating hardware in networks. Such type of computing, by which computing services are not delivered independently but dependent of networking services, is named Aqua Computing. The authors see Aqua Computing as a novel approach for delivering computing resources to end devices, where computing power of the devices are enhanced automatically once they are connected to an Aqua Computing enabled network. The process of resource coupling is named computation dissolving. Then, an Aqua Computing architecture is proposed for mobile edge networks, in which computing and wireless networking resources are allocated jointly or cooperatively by a Mobile Cloud Controller, for the benefit of the end-users and/or for the benefit of the service providers. Finally, a working prototype of the system is shown and the gathered results show the performance of the Aqua Computing prototype.
\end{abstract}

% not want either math or citations to appear in the abstract.

}

\maketitle

%\tableofcontents

% is not selected <OR> if conference mode is selected - because compsoc
% conference papers position the abstract like regular (non-compsoc)
\IEEEdisplaynotcompsoctitleabstractindextext

% \fi
\IEEEpeerreviewmaketitle

\section{Rational and Objective}

\IEEEPARstart{W}{hen} the term Ubiquitous Computing first coined by Mark Weiser \cite{Weiser:1999:CSC:329124.329126}, in his vision future computers would come in different sizes and will be pervasive in humans' life by appearing everywhere to activate the connected world. Furthermore, people will unconsciously use them to perform everyday tasks. Computers virtually and metaphorically will be invisible to the user, or they will be vanished into the background, so that the users are freed to use them without thinking. As of today, one would argue that Weiser's vision has somewhat become a reality. However, we believe that there is still a long way to go for computing to be truly pervasive. Cloud computing provides ubiquity, on-demand resource allocation, scalability, to computing. With the advancements of smart devices, Mobile Cloud Computing (MCC) combines cloud computing technologies and mobile computing. MCC uses availability and elasticity of cloud computing resources to provide seamless executions of Rich Mobile Applications (RMA). One of the objectives of MCC is to enhance user experience with unrestricted computing and network functionalities.
 
While, MCC investigates into mobile computing rather from the mobile devices' standpoint, Fog Computing \cite{bar2013fog} and Edge Computing (MEC) \cite{patel2014mobile} bring computational resources to the highly distributed edge of the network from centralised data centres. Bringing computational resources to a closer proximity to the end-users results in local resource pooling, increase in redundancy and improved Quality of Service (QoS). Another aspect of fog computing is Internet of Things (IoT) support, where most of the devices that we use today are connected to each other for providing better user experience. The mobile devices get access to fog resources using existing cellular or WiFi networks.
 
 \begin{figure*}[ht]
 \centering
 \includegraphics[]{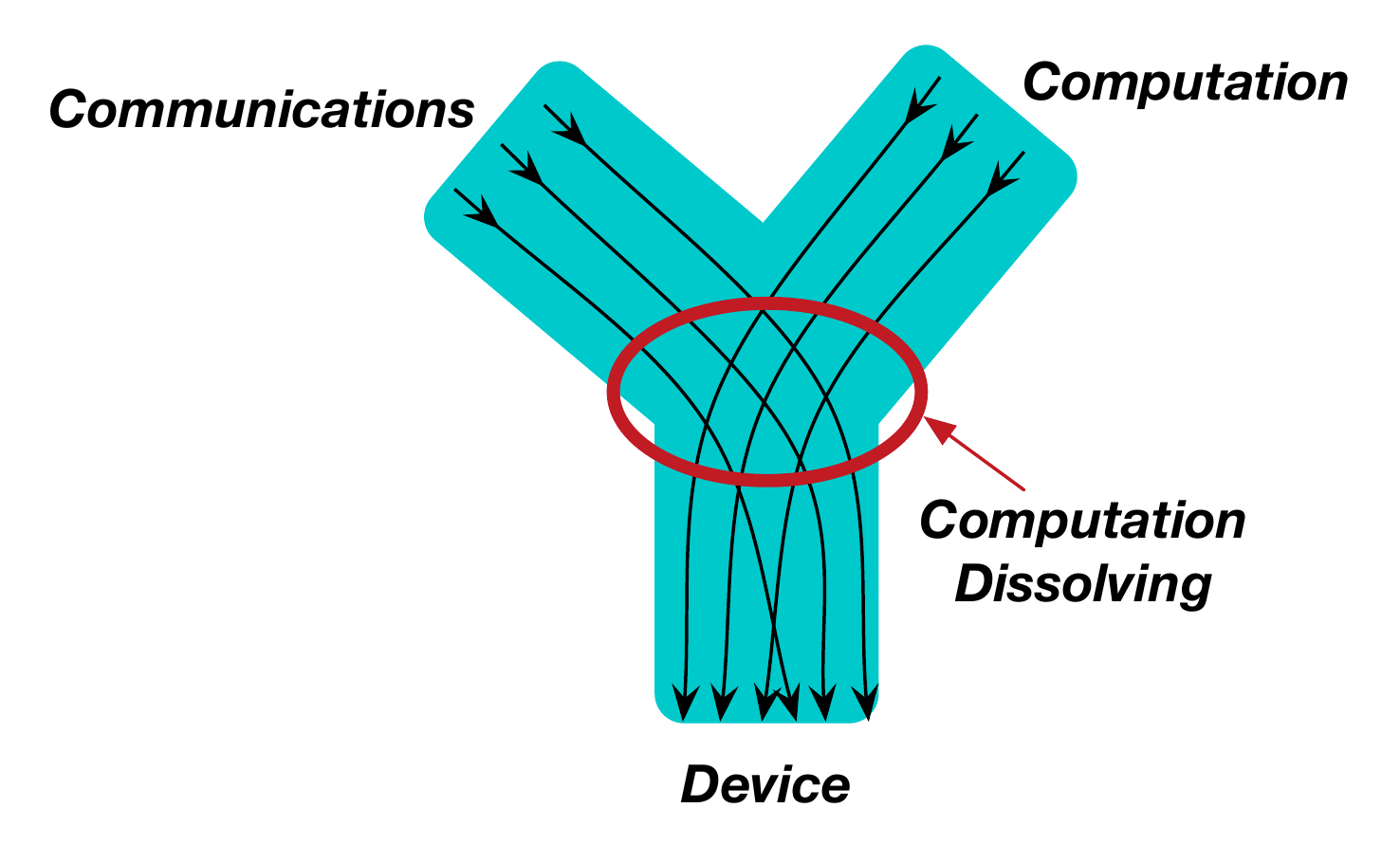}
 \caption{\label{fig:discomp} Computation Dissolving: Coupling Computing and Communications}
 \end{figure*} 
 
Although, all aforementioned technologies have separate data and control planes, management of all associated resources are carried out independently. From the end-user's point of view the received resources from service providers, communicational and computational resources are delivered independent of each other. We see a need for a coupling between resources for providing better optimised resources for the end-users and to better optimise providers' infrastructures. An existing example for resource coupling is, power over Wi-Fi \cite{DBLP:journals/corr/TallaKRNGS15} that delivers power to connected wireless devices coupled with network resources to connected devices. Once the users are connected to the network, the batteries of the devices can be automatically recharged. The authors envision the future of computing involves coupling various computing and communication technologies, instead of providing independent resources or services to the end users. For example, today the end-users receive network bandwidth from a mobile network operator or from a fixed network service provider. To expand computational or storage resources they have to seek for separate service providers. Although, in future new types of resources will emerge, where they couple two or more conventional resources together. Such coupling will be done distributed throughout the networks to achieve common objectives, such as to reduce energy consumption of the mobile devices and of the infrastructure, increase Quality of Experience (QoE) or to introduce new types of services. In future, the users will pay the network operators not only for the network bandwidth and the consumed data, but also for computing power that it consumed for powering up their devices. 

In present-day, the manufacturers try to make the mobile devices thinner by reducing the size of its hardware components. Nonetheless, we envision that computation-communication coupling will take some or most computing away from the end devices, while also bringing computing closer at the same time, by handing over computing tasks from devices to the network.
 
The tight coupling of computing and communication resources make computing ubiquitous. The computing resources that are made available to end users by the network allows the connected devices to be less equipped with computing resources and migrate computing tasks to the network. Consequently, in a ubiquitous computing environment the connected devices will become thinner. Furthermore, compute-communication coupling enables the implementations of smaller devices that were difficult to implement due to their size constraints and high computational demands. From the end-users' perspective, such coupling helps computing to disappear into the background among other resources such that, users receive computing services when connected to a network without needing to explicitly interact with the system. Once the computing and storage resources are acquired by the users, to perform computing tasks the users may subscribe to other services or will download applications that will utilise computing resources that are deployed in the network.

The authors name this novel approach to computing, Aqua Computing, a paradigm that closely couples computing and communications together. The name Aqua is inspired by the water cycle, and also known as hydrologic cycle. Where, water can take any form of a cloud or a fog. Likewise, referring back to Cloud Computing, we see Aqua (water) as the essence of computing as it may take any form of a Cloud, a Cloudlet or a Fog. Likewise, relating to the location, water can be in clouds (central data centres), in fogs (at the edge of the network) or on earth as liquid (in devices).  The process of coupling computing resources with communication resources is seen as dissolving computational resources, as shown in Figure \ref{fig:discomp}. 
 
Aqua computing enables the device manufacturers migrate hardware and software functionalities of mobile devices to the connected networks and their clouds. Furthermore, hardware functions will be replaced by software modules that are deployed in Aqua Computing environments. Consequently, the size of mobile devices become even thinner, allowing the mobile vendors to remove hardware modules that are replaced by software components in clouds/networks. Subsequently, one may envision future devices being thin and compact devices that will just have display and network capabilities, while the computing tasks will be done distributed in networks.

In section \ref{sec:cdissolving}, the authors discuss computation dissolving methodologies. Then in Section \ref{sec:aquabenefits}, the benefits of aqua computing for both end-users and service providers are explained. A new architecture for aqua computing is proposed, in Section \ref{sec:theclones}, and a developed prototype for further evaluating the Aqua Computing concept, in Section \ref{sec:mcprototype}.
 
\subsection{Computation Dissolving}
\label{sec:cdissolving}

\begin{figure*}[ht]
 \centering
 \includegraphics[]{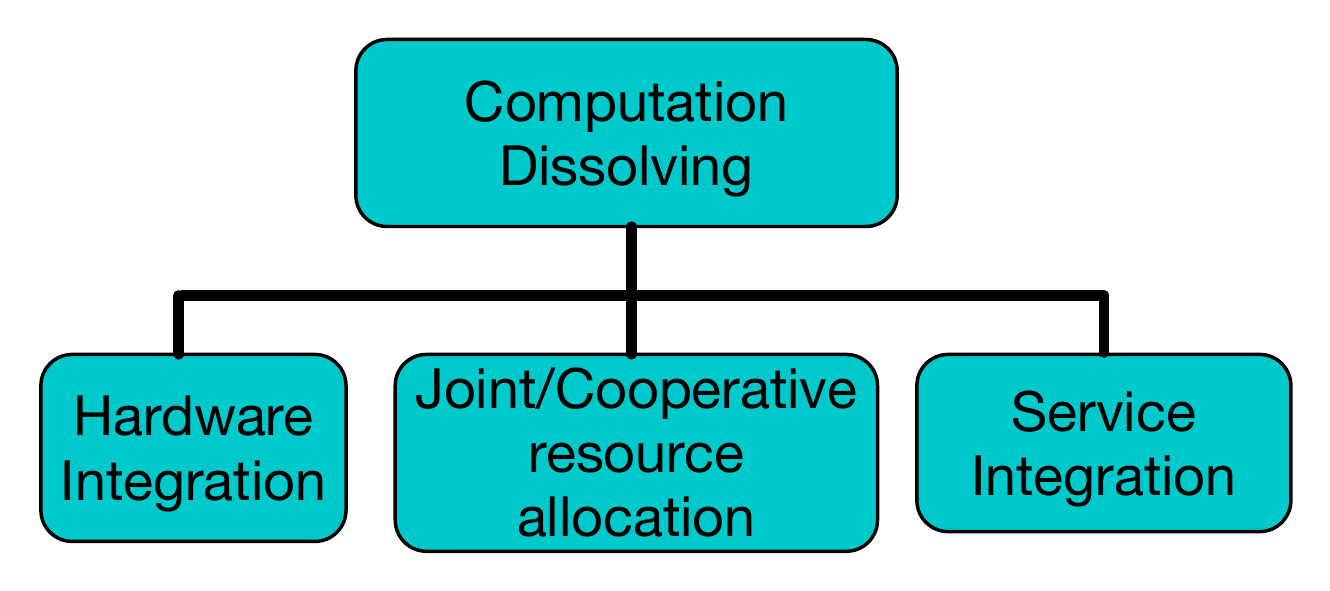}
 \caption{\label{fig:dismethods} Computation Dissolving: Methodologies}
 \end{figure*}

 For making end devices computationally powerful, computation dissolving is done by future service providers. Various methodologies will be used for computation dissolving as shown in Figure \ref{fig:dismethods}.

\subsection*{Joint/Cooperative resource allocation}

Resources which may or may not be in the same domain may be jointly allocated, for providing better services or to optimise resources for achieving high efficiency. For example, joint resource allocation between computing and communication domains. In such environments, connected devices are much more powerful than the ones that are not. In aqua computing, at a given time the computational resources are delivered to the end users without they actively requesting for it. For dissolving computational resources with communication resources in Radio Access Networks (RANs), wireless baseband resources can be jointly or cooperatively allocated with mobile edge computing resources to the end users for providing seamless services while also optimising networking resources. Similarly to wireless networks, broadband resources in fixed networks will be allocated jointly or cooperatively. Cross layer optimisation between the layers of the OSI communication model and computing layers, can increase pervasiveness. Examples for layers in the computing domain can be seen as, software applications, virtualisation layers, middleware, operating systems.
 
\subsection*{Hardware Integration}

The authors envision in future computing, computing hardware integrating with hardware components that perform tasks on other types of resources.  Multi purpose hardware devices will be designed and implemented to perform tasks with different types of resources at the same time, for providing computational resources to the end-users. Seemingly, computing pools will be integrated with communication devices. For example, computing cloud integration with base stations, access points, broadband cabinets, telecommunication exchanges, gateways, for delivering services that couple networking services and computing services together. 

\subsection*{Service Integration}
 
The authors believe, to achieve true pervasiveness, communication services, compute services and storage services has to be integrated together. In future, end-users will not subscribe to individual services separately, but to a one service that is provided by coupling multiple services in various domains. Such domains may include, communication, computation and storage.

\section{Benefits}
\label{sec:aquabenefits}
\subsection{Benefits for the User}

There are multiple aspects of Aqua computing that benefit the end-user. Due to computation dissolving, the users do not have to request or subscribe to multiple services at the same time. For example, a mobile network operator may provide computational resources with its mobile network service for the subscribers.

The users may observe that their mobile devices can perform computational tasks much faster when it is connected, than when it was not. Moreover, the mobile applications and the operating systems may become faster and efficient while also increasing QoE. The application developers will take network parameters into account when developing applications for devices in Aqua Computing environments. For executing applications in an Aqua Computing environment, network information will be very important. Henceforth, the applications will be aware of the network parameters such as bandwidth, delay signal strength, the location in the network that computational resources are placed in and the size of the computational resource.

As a case in point, one may imagine that there are two Mobile Network Operators(MNO), which have the same network bandwidth and latency when accessing the internet, called A and B. The only difference between the two is, A is a conventional MNO and does not have Aqua Computing deployed, but B has Aqua Computing technologies have deployed. If there is a mobile user that can connect to both networks, when the user connects to A, he is able to get access to the internet just like any other network that exists in present day. Although, when the user connects to B, he is not only able to surf the internet faster, but in the background the computing capacity of his mobile device has also elevated while improving the user experience. Such increase of computing capacity of the connected devices will not only help perform networking operations even faster, but most importantly it will also increase the performance of the mobile operating systems and other installed applications. This is a result of computation dissolving that helps to achieve computing pervasiveness. 

Aqua Computing will also make the mobile devices thinner and lighter by converting hardware functions of devices to software functions, which runs on computing resources in the network. Transferring computing tasks to the network may also reduce energy consumption in the device. A similar phenomenon has started happening today where power charging technology is allowing electric cars to be charged wirelessly, as they are being driven on roads. By allowing cars to wirelessly charge, the size of the battery that a car needs to carry may reduce, making future electric cars lighter.

\subsection{Benefits for the Service Provider}

Network and computational resources will be jointly and cooperatively allocated to optimise the performance of networks and services. For an instance, one approach is to jointly allocate computation and network resources, while exploiting computing and network transmission trade-off shown in Equation \ref{eq:offtradeoff}, when computational tasks are offloaded to the network \cite{5445167} . Where, $T$ is the time takes to execute a task, while $D-S$ amount of data is transferred with a $r$ bitrate to the cloud, for executing $F$ number of instructions in $f$ speed. The set of data $D$ is required to execute the given task, while some parts of that data $S$ might already exist in the network. The objective of the resource allocation algorithm may vary, more computational resources $f$ can be allocated for the benefit of the network or more network resources $r$ for the benefit of the computational resources in the network.

\begin{equation}
\label{eq:offtradeoff}
T = \frac{D-S}{r} + \frac{F}{f}
\end{equation}

The introduced computational resources in the network may increase the quality of the resources and services that are offered by the service providers. The quality of network resources will also be increased as computation dissolving increases QoS. Aqua Computing integrates computing services with other services including network services.

Accordingly, if Aqua Computing is deployed by a MNO, the subscribers' QoE will be increased, as a result of computation dissolving in the network. Subscribers will notice a significant improvement, in not only the network performance but also application performance of their devices locally, while they are connected to the Aqua Computing enabled MNO's network. Online gaming is another application where Aqua Computing can help enhance users' experience. Enabling Aqua computing in the fixed network will allow fixed network service providers, to provide gaming services to the send users. In such scenarios, Aqua Computing enables network service providers to offer computing rich new services to the end users along with network services.

Computation dissolving may introduce new types of services. Such improved services and new types of services may lead to new price plans for the users. There, also will be a change in the current business model of the network service providers. Instead of the subscribers having to subscribe for compute services from different service providers, the computational resources that are built into the network will allow the network service provides also to offer compute services to the end-users. All aforementioned, will lead to better price plans that increase revenue for the service providers while also increasing the user experience for the end users that may attract more customers.

\section{The Clones}
\label{sec:theclones}

\begin{figure*}[ht]
\centering
\includegraphics[width=\textwidth]{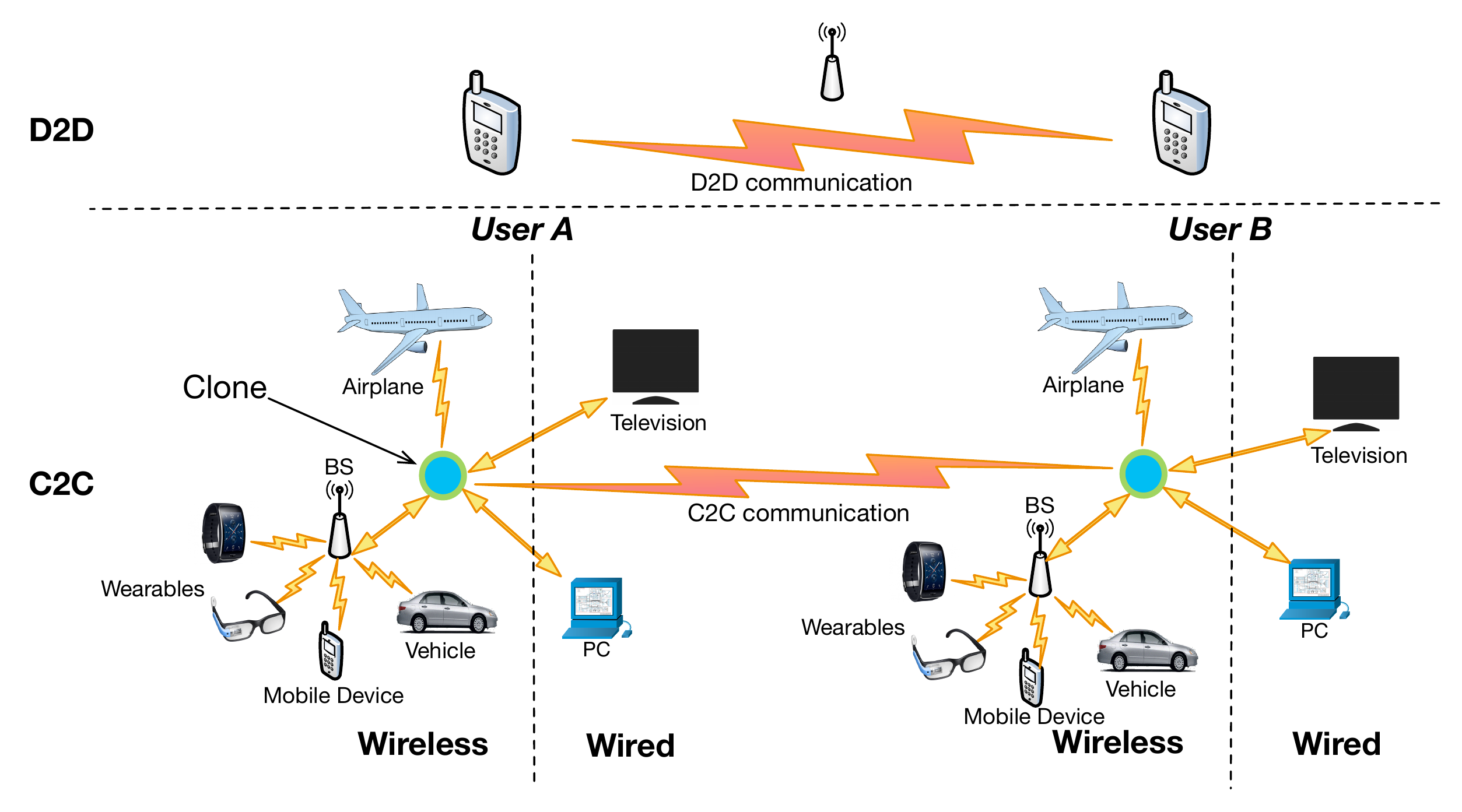}
\caption{\label{fig:ucomp} Clone: an architecture for Aqua Computing}
\end{figure*}

The clone is the main element of the proposed architecture for Aqua Computing. The clones are user specific, i.e.\ all end-users may have their own dedicated computing spaces in the network; one clone per user. They are placed in the network and are also managed by the network service providers. One speciality of the clone is, in networks each subscriber has their own clone, as a segregated computing space from the rest of the users. Secondly, it is not only a computational space, but also a space for storage. The nature of the clone is, one clone is dedicated to only one user at all times. The clone perform tasks, make actions, made decisions, store data behalf of its user. Therefore, we believe that the clones can also be seen as virtual personalities of the users, for the rest of the network and for the internet. The clone may also be available to the end users via wireless and wired networks as shown in Figure \ref{fig:ucomp}. Furthermore, some of intelligence and computational tasks that conventionally executed at the end user's end can be migrated to the clone. 

\subsection{Computing resources}

Computational resources that reside in clone can be dynamically allocated to the user by the resource allocation algorithms. The computational resources can be shared between multiple applications and connected devices. One of the most important benefits of the clone is, it lets other devices offload compute tasks to be executed in the clone. As depicted in Figure \ref{fig:ucomp}, one user may have multiple devices, and the clone becomes a central point where all devices of the same user synchronises, allowing the clone to manage all devices centrally behalf of the user.

The computing resources of the clone can also be used for making decisions behalf of the user. The authors see this as, offloading intelligence from devices to the clone. The clone will carry out context-aware decisions making tasks in the clone, with the help of the network information, application information and information from user data that it has access to. When the clone is at the edge of the network, for computation dissolving the clone has access to network information from the base station or the BBU pool that the corresponding user is connected to. Since the user's devices are attached to his clone, the clone is aware of all the applications and devices that user uses, and the user might use the clone to expand storage of his devices by storing user data in the clone.

Cloud computing is a model that has been widely used for delivering computational resources. In present day, cloud computing offers virtualised resources of hardware resources to the end-users. Virtualisation is carried out in various levels in computing. Similarly, a clone is a virtualised computational resources, such as a virtual machine that offers computational resources to the connected devices.

\subsection{Communication resources}

Clones may use network sockets for making connections to the nodes in the internet and to other clones in the network. Dynamic resource allocation algorithms in Compute dissolving process will dynamically allocate communication resources such as bandwidth, network sockets and bit streams for the clones. The network infrastructure of the MNO will be used by clones to communicate between clones and to communicate  with the networks outside the MNO's network (Internet). 

As mentioned in previous sections, the clone posses storage resources for storing user data. This storage may also be used for data caching to make communication efficient. The clones will cache frequently accessed information, application code and other user data for reducing network traffic of, device-to-clone (fronthaul), clone-to-clone and clone-to-internet (backhaul) networks.

\subsection{Computation scenarios}

Computational resources that are in the clone is primarily used for Compute offloading, where chucks of computing code are transferred to the clone, and once executed the results are sent back to the mobile device. Computation offloading is a methodology that is used for increasing the computational capacity, and to increase energy efficiency of mobile devices. The mobile devices may offload computing tasks to the clone using wireless networks, while other devices that are connected to the wired network is able to offload tasks to the clone using the wired network. For computation dissolving, a dynamic resource allocation algorithm that resides in a centralised controller will allocate bandwidth for transferring code to and from the clone, and the amount of computing resources in the clone for executing the received computing tasks.

The mobile device may offload its decision making tasks to the clone, taking accurate network information that it is able to acquire from the network, its context-awareness and computing capabilities. For an instance, adaptive bitrate streaming is a widely used technique that is used for streaming multimedia over networks. The streaming server makes the same content in multiple bitrates available for the requesting user to choose. Periodically, the mobile device switches streams depending on the available network and CPU resources, for increasing QoE for the end user, i.e.\, decision making for adaptive streaming is done in the mobile device. Considering, the placement of the clone in the network, and its ability to acquire information of the connection between its dedicated mobile device and the base station, we can offload decision making of adaptive streaming to the clone. Whereby, instead of the mobile device, the clone detects the network connectivity status of the mobile device from the base station, then decides and instructs the streaming server to switch between streams of different bitrates. Aforementioned, will reduce computing demand and power consumption of mobile devices when streaming video.

\subsection{Communication scenarios}

\begin{figure*}[ht]
\centering
\includegraphics[scale=0.7]{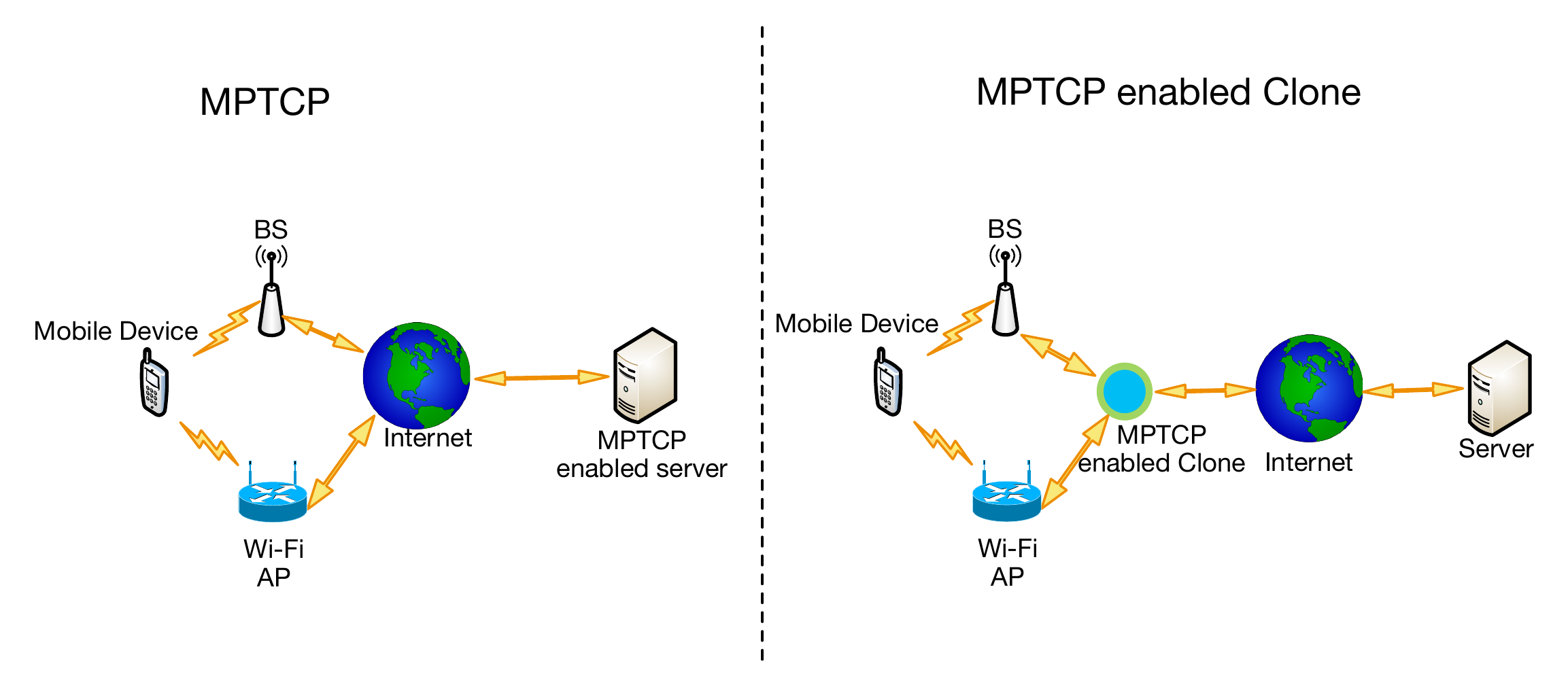}
\caption{\label{fig:cloneproxy} Clone as a MTCP proxy for seamless service delivery}
\end{figure*}

The clones may help for communication related operations of the mobile devices in various means. Multi-path TCP (MPTCP) enables devices to initiate more than one TCP flow to the destination that it communicates with for increasing TCP throughput. As of today, MPTCP has started to become available among consumer devices. In heterogeneous networks MPTCP is used for enabling seamless hand overs, as well as for increasing throughput. While the clone is at the edge of the network, one can use MPTCP enabled clones for assisting MPTCP operations of mobile devices in wireless networks. One quirk of MPTCP is, both communicating parties has to support MPTCP protocol. However, when communicating with nodes in the internet, one can not be certain if MPTCP would be enabled on all devices that they communicate with. However, the clones can help with this matter. In the left side of the Figure \ref{fig:cloneproxy}, shows the conventional MPTCP scenario where a mobile device accesses a MPTCP enabled remote server using both cellular and Wi-Fi networks. As shown in the right side of the figure, the MPTCP enabled clone acts as a MPTCP proxy terminating the protocol at the clone. In this scenario, the nodes in the internet that the mobile device communicates with does not have to support MPTCP, as MPTCP terminates in the clone. The network connection from the clone to the internet is a conventional TCP connection.

There are several benefits of sending network traffic of mobile devices through their clones. In a video streaming scenario, one of the benefits of sending traffic through the clone is that, the clone is able to act as a buffer point at the edge of the network for the video stream. For example, one aspect of it is, considering the instabilities of the wireless network, the introduced buffer point can help reduce retransmission traffic in the backhaul network of the MNO, in a RTMP (Real Time Messaging Protocol) video streaming scenario. RTMP being a TCP based protocol, TCP's reliable stream delivery service retransmits lost and damaged packets back to the client. In a traditional system, the retransmission is done by the source of the content, i.e\ the video streaming server. However, the clone acting as a buffer point in the middle of the source and the receiver, it is now able to do retransmission of packets, instead of the video streaming server, given the buffer contains the retransmitting packets. One can see this as, the clone hiding wireless related communication issues from the fixed network.

Another aspect of clone communication\cite{kosta2012clone2clone} is, turning Device-to-Device (D2D) communications into clone-to-clone communications, as depicted in Figure \ref{fig:ucomp}. These communication scenarios of the clones are twofold. 1) Device-to-device communication between the devices of the same user, 2) and device-to-device communication between the devices of different users, offloaded to the clone. One can see this as converting device-to-device communications into clone-to-clone (C2C) communications. In the first scenario, due to the fact that all devices of the user is connected to the same clone, allows one device to communicate with the rest of the connected devices, through the clone. This eliminates the use of wireless networks for transferring data between devices. The data caching in clone makes C2C communications efficient. 

Device to device communications enables direct communication between two devices, in the same vicinity, using existing cellular spectrum. However,  content sharing can be made efficient for the network provider with a clone in place, while extending storage capacity for the mobile user. Users share content using various applications and using various platforms (e.g.\ Email, Whatsapp, Viber, WeChat, iMessage, Facebook, Instagram). Since, all users have their own clones, when one has to send data to another, the sender's clone can transfer data to the receivers clone assuming that sender has already got the requested data in the clone. These kind of communications can be seen as Clone-to-Clone  communications. In future mobile social networks, users store their User-Generated Content (UGC) on their clones rather than on a centralised server (e.g. Facebook, Instagram servers). Future social networks enabled by clones store content on content owners' clones. When sharing content, C2C communication is used, rather than sending data all the way to other public cloud servers and back to the receiver. We consider following C2C scenarios.

% Sort out subsection issue. Here, you have to make sure C2C scenarios are in a sebsection, not in the same level.

\subsubsection*{One sender - One receiver}

\begin{figure*}[ht]
\centering
\includegraphics[]{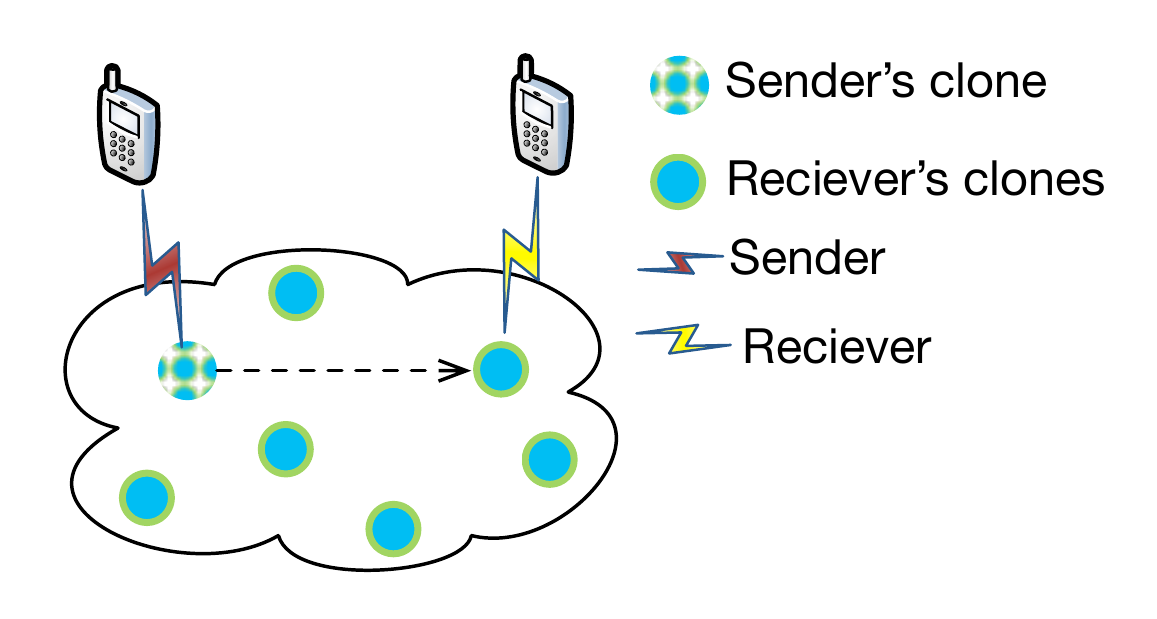}
\caption{\label{fig:c2c1} C2C: One sender - One receiver}
\end{figure*} 

In this scenario, user A (sender) may upload its data that is intended to be shared with others in the future, to its dedicated clone. User B (receiver) is interested in what user A owns, so he requests user A's content. Once, a connection has been established between the user A and B for sharing requested data, user A sends the requested content to user B's clone. Finally, the user B receives the requested content from its own clone. In this scenario, the users or the network operators may not benefit. Since, similarly to the D2D communication scenario the same amount of wireless spectrum has been used ones to upload (Sender-to-clone A) the data and also again to download the data at the receiving end (Clone B-to-receiver). It is only when the sender needs to send the same content more than one time, the cellular spectrum can be saved due to data caching that is done in clones. Since, the receiver's clone can get the content from the senders clone without the sender having to re-upload content to his clone more than one time, assuming that the sending clone has still got all content stored when the second receiving user requests for content, as shown in Figure \ref{fig:c2c1}.

\subsubsection*{One sender - many receives same content to their own clone at the same time.}

\begin{figure*}[ht]
\centering
\includegraphics[]{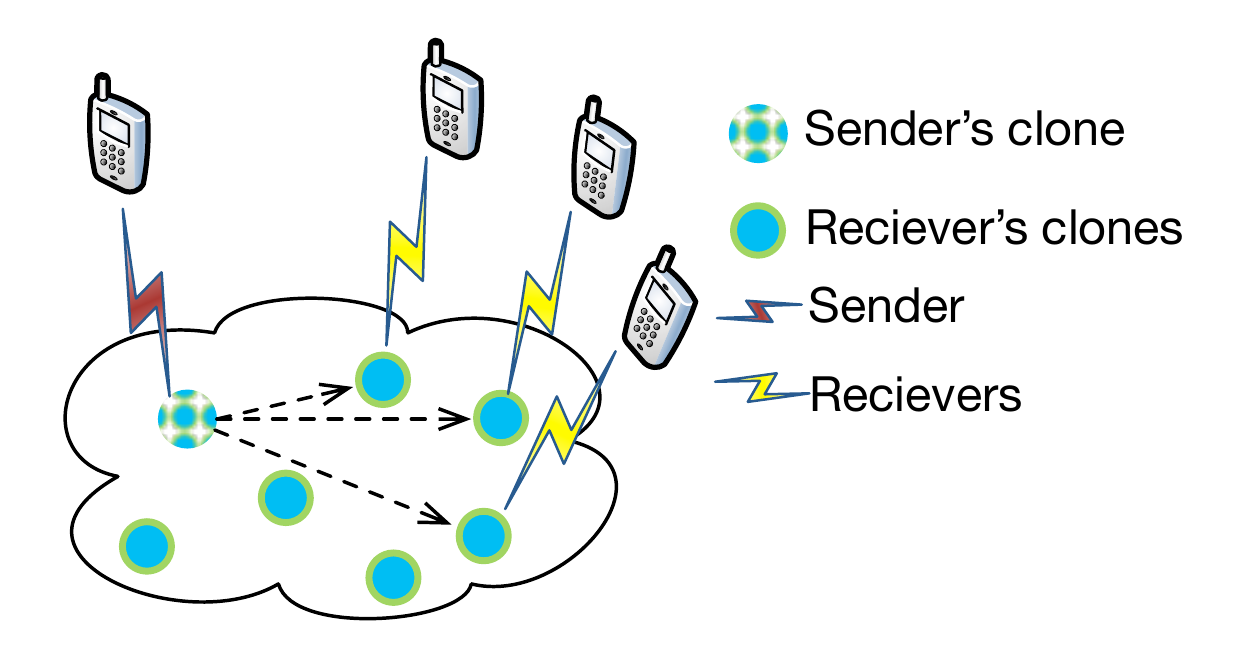}
\caption{\label{fig:c2c2} C2C: One sender - many receives same content to their own clone at the same time.}
\end{figure*} 

A (sender) has got something $n$ users are interested in. All interested users request for the same content from the user A. Once, the receivers have established connections with user A's clone to receive data. The sender uploads the content to his clone. Then, the sending clone sends requested data to all clones of the receivers. In this scenario, wireless spectrum can be saved since the sender only uploads the content ones when sharing the same content with many users at the same time. Whereas, in a conventional D2D scenario, the sender has to upload requested data $n$ times when sending data to all $n$ receivers, as depicted in Figure \ref{fig:c2c2}.

\subsubsection*{One sender - many receives same content from the sender's clone directly.}

\begin{figure*}[ht]
\centering
\includegraphics[]{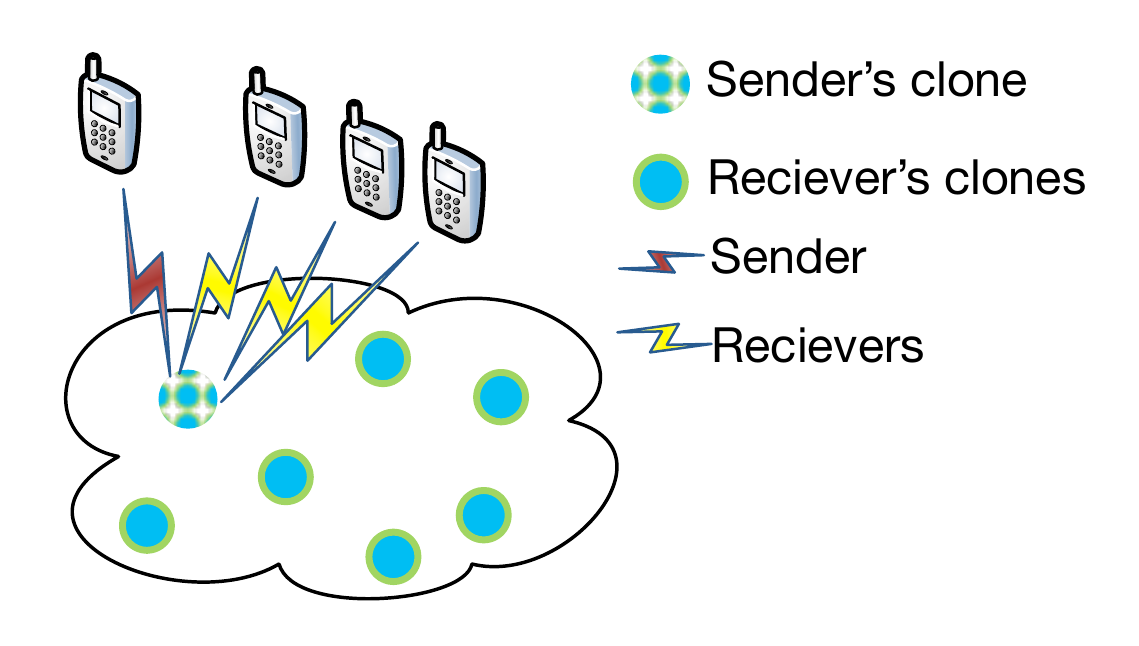}
\caption{\label{fig:c2c3} C2C: One sender - many receives same content from the senders's clone directly.}
\end{figure*} 

The scenario that is shown in Figure \ref{fig:c2c3}, is quite similar to the previous scenario in Figure \ref{fig:c2c2}, but the only deference is, the sender does not have to send the content to receivers clones. The senders clone allows all other receivers to get content directly from his clone. This scenario eliminates the second C2C data transfer step. The benefit of this elimination is that, network traffic reduction that is achieved in the mobile backhaul network.

\subsection{System architecture}
\label{sec:mcarch}

The clones are highly distributed computing elements, that are deployed for the benefit of both end users and the mobile network. Due to the fact that the operating system that runs on the mobile clone has to be the same operating system that runs on the corresponding mobile device for compute offloading, it is assumed that clone versions of mobile operating systems are made available for operators to deploy.

Providing a solid infrastructure for delivering network services to the end users has been the main purpose of the mobile network operators today. The clones open up ways the operators can help to improve the user experience and the services that are delivered to the users, as well as making the content delivery faster and reliable. Furthermore, operators are interested providing additional services and performance enhancements on top of their mobile network services. The clone enabled network operators may become more than just network pipe providers. The clone is a platform that, also enables other service providers to build up services on top of it, i.e.\ cloud service providers and application developers may use clones to enhance performance of their services.

\begin{figure}
\centering
\includegraphics[width=\textwidth]{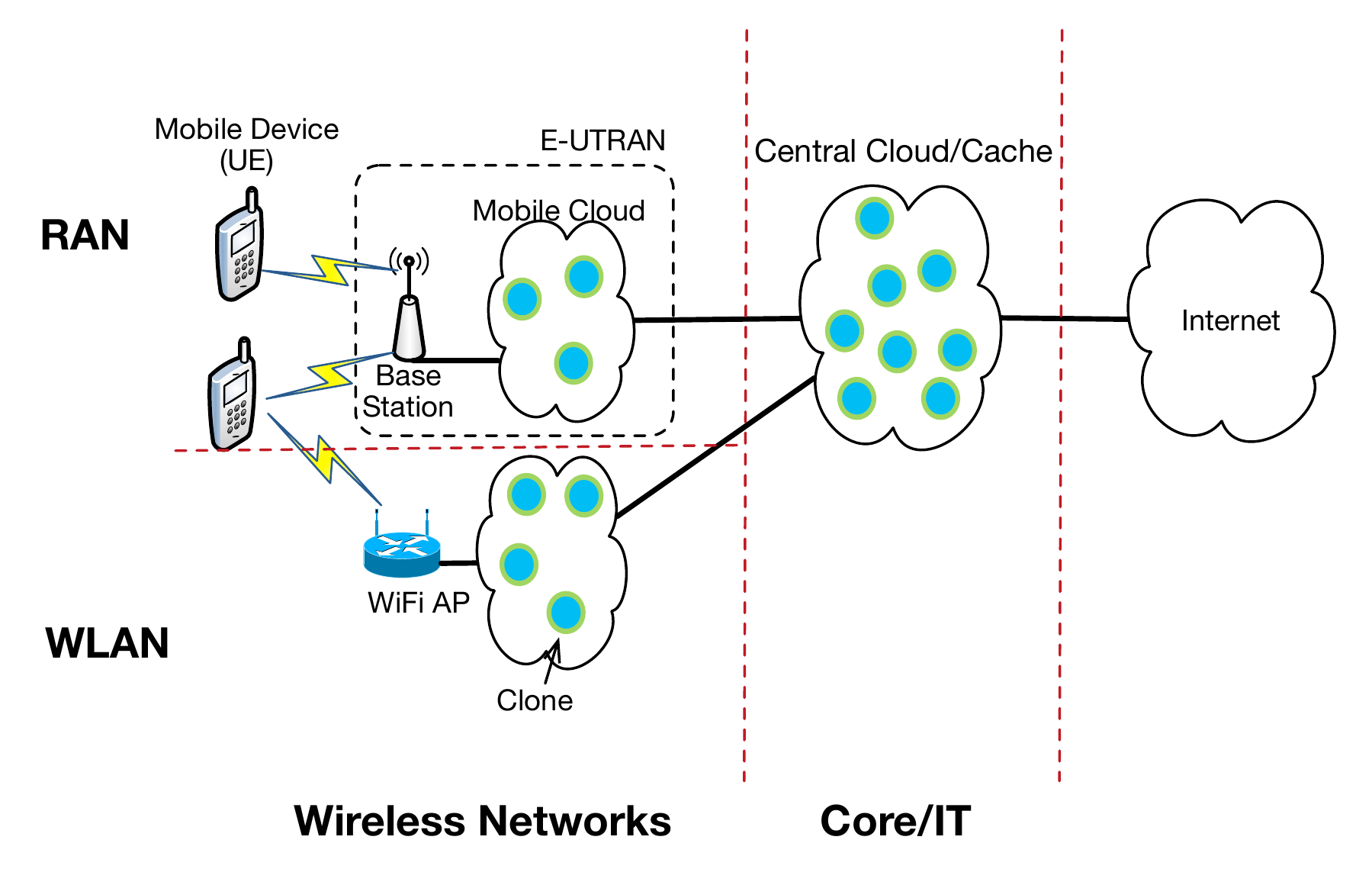}
\caption{\label{fig:mcarch} Mobile Cloud in RAN architecture}
\end{figure} 

There have been a vast amount of work has been done on resource allocation wireless network over the years, to allocate wireless resources among the users. Similarly, cloud computing have evolved largely for the past few years, and there have been significant improvements in virtual and physical resource allocation in data centres.
The authors introduce a cloud of clones, a mobile cloud into the wireless networks, as depicted in Figure \ref{fig:mcarch}. The additional computational resources put alongside the edge network resources, has lead researchers to rethink how computing and network resources should be allocated together \cite{6923537}. This approach of allocating resources in computing, is named computation dissolving. Furthermore, a clone may also maintain user subscription information and the subscriber's identity (Electronic Subscriber Identity Module [e-SIM]), so that the clone cloud is able to associate the clones with their corresponding users and also able to tailor the clone services according their subscription price plan. i.e the clone size and the hosted services in clone may vary depending on the type of mobile subscription. Additionally, there may be cases where all inbound and outbound traffic are routed through the clone for adding value added service on the clone.

\begin{figure*}[ht]
\centering
\includegraphics[width=\textwidth]{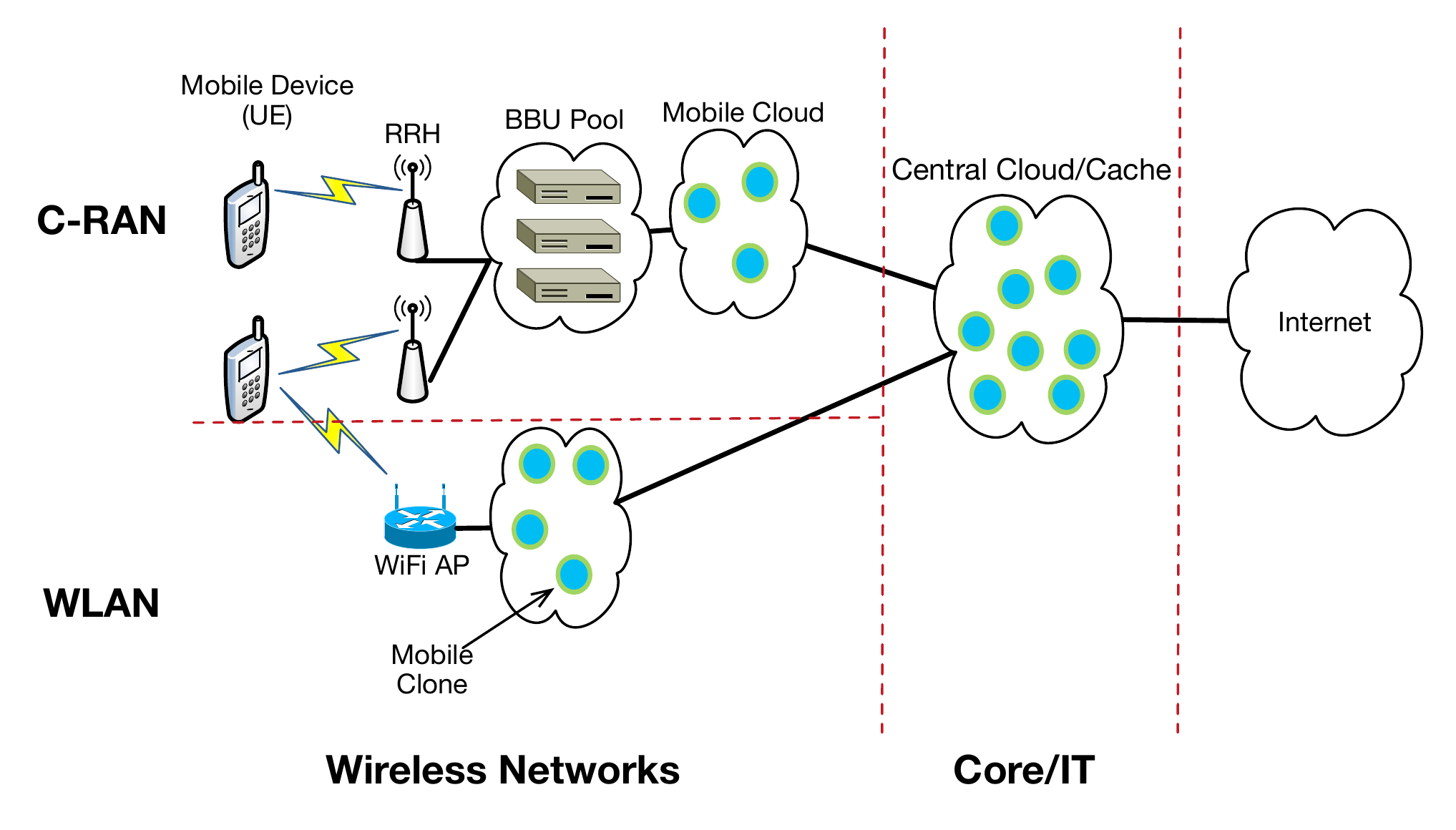}
\caption{\label{fig:cranmcarch} Mobile Cloud in C-RAN architecture}
\end{figure*}

Cloud-RAN \cite{chenc} is a new cellular network architecture for future mobile networks. For a C-RAN specific approach, the mobile cloud is placed next to the pool of Base Band Units (BBU) as shown in Figure \ref{fig:cranmcarch}, where the Remote Radio Head(RRH) units are deployed separate from the BBU pool. One of the design goals of such an approaches is to enable joint and cooperative resource management between the BS/BBUs and the mobile cloud \cite{ikwangj:2015TOC}. As the mobile cloud is placed in the operator's network makes communication and computing operations easier. 

The mobile clouds may also be placed next to WiFi access points. The benefits of this approach is that, it can take advantage of the higher bandwidth of WiFi networks. There are two deployment scenarios; 1)  mobile cloud with home Wifi APs 2) mobile cloud next to public APs. In the first scenario, the amount of users are static, so that a fixed number of clones can be assigned to the known users in the household. In the second scenario, the number of connected users are dynamic.

There are two types of cloud resources in both Figure \ref{fig:mcarch} and \ref{fig:cranmcarch}. Persistent cloud in Central Cloud (Core/IT) network and transient cloud resources at the edge of wireless networks (Mobile Cloud). Persistent cloud resources continue to exist and serve their dedicated mobile devices. They do not only contain computing resources, but also provides storage services to the mobile users, where they can store data such as, user-generated content, cached data and other data that belongs to their dedicated mobile users. Hence, it is of type persistent. On the contrary, as the name suggests transient cloud resources are temporary computing resources that are placed at the edge of the networks. The mobile cloud resources at the edge of the wireless network are made temporary due to following considerations 1)  as the clones are user-specific, to keep the clones at the closer edge network, as they move geographically the clones can also be efficiently migrated close to users. 2) due to space and size limitations of resources that can be placed alongside every BS or BBU pool of the mobile network. Such that, once a mobile device is connected, a clone is spawned in the mobile cloud at he edge of the network and all other relevant user data are pulled down to the edge of the network from the Core/IT network. Subsequently once the tasks between the mobile device and the clone has been terminated, the clone can be destroyed to make space for another user. The Central Cloud is a data centre type cloud environment, where the MNO maintains all persistent clones of all subscribers of the network. A central cloud can also be regional, where it only hosts the clones of the subscribers that happen to reside in its region. Regional central clone clouds may have connectivity between other regional clouds to support clone mobility. 

For aforementioned edge and central cloud operations and communications, intercloud \cite{5072540} and intra-cloud protocols need to be designed. Requirements for such protocols are twofold, 1) intercloud protocols to realise interoperability and federation between edge and central computing resources. 2) intracloud protocols, for federating computing and communication resources within edge or local clouds in networks.

At this point in time, virtual machines have been used as the primary technology for building cloud service infrastructures. Although, recently there are other virtualisation technologies that have started become popular, such as Linux Containers (LXC). There is a need for better virtualisation technologies for mobile clouds that are optimised for high delay sensitivity in mobile network operations. 

Mobile compute offloading is one of the primary objectives of the mobile clones. In conventional mobile compute offloading systems, the mobile network is not aware of the offloading process. The network treats the offloading code and data as any other data. However, computation dissolving makes the network aware of the compute offloading code. Specifically, mobile cloud controller makes the network aware of the computing resources. Secondly, conventionally the compute offloading destination resides outside the network operators' network. The offloading process is made efficient by dynamically allocating resources to satisfy offloading requirements.

\subsubsection*{Hardware Integration: Evolved mobile Base station}

\begin{figure*}[ht]
\centering
\includegraphics[scale=0.7]{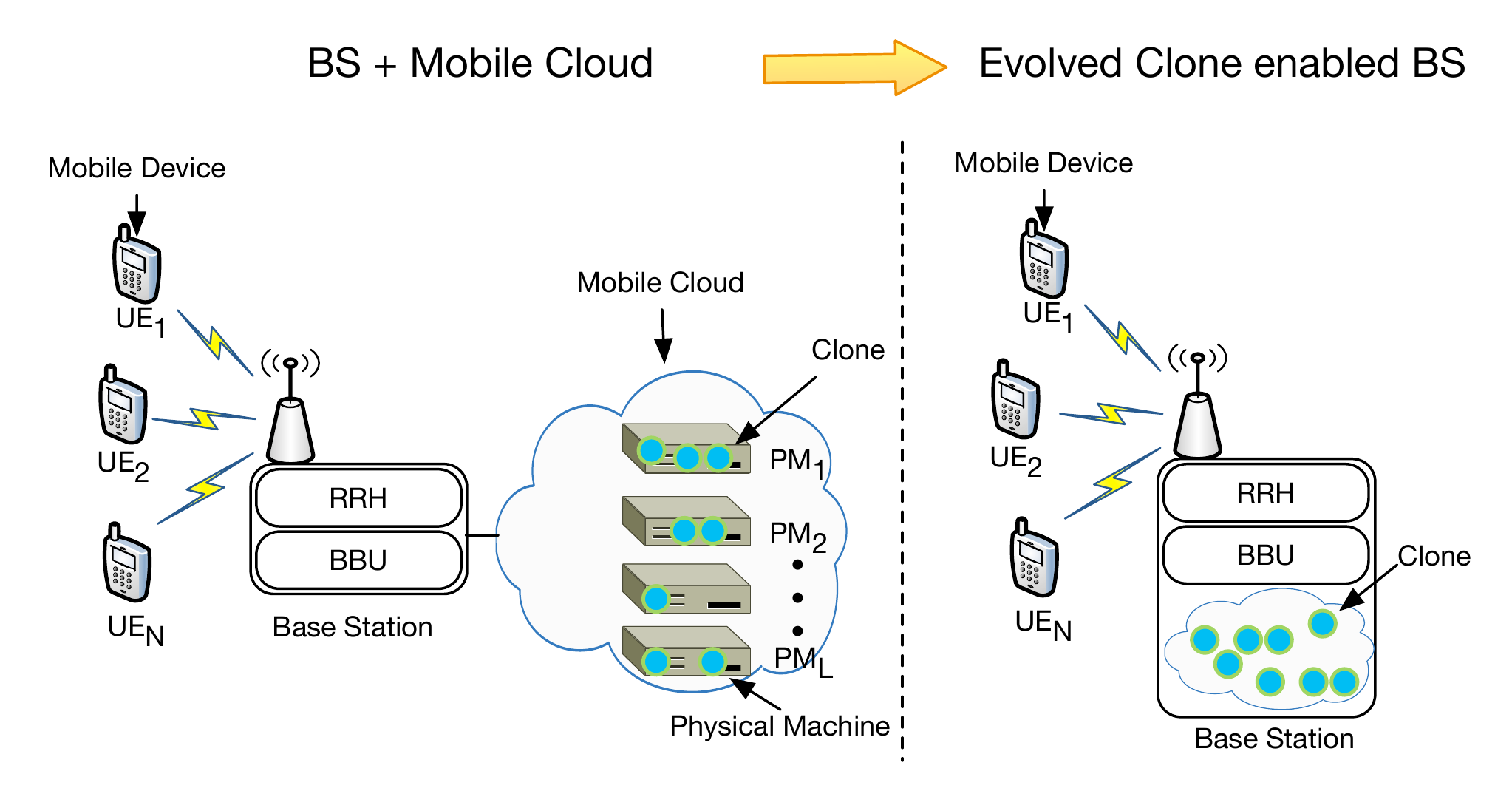}
\caption{\label{fig:evolvedbs} Computation Dissolving: Evolution of Clone enabled BS}
\end{figure*}

Throughout this paper, the authors treat the mobile cloud as a separate entity from the base station or from the BBU pool. However, the authors foresee these two entities merging into a one entity as shown in Figure \ref{fig:evolvedbs}, as the technologies advance in time, exploiting computation dissolving in hardware level. 

The left side of the Figure \ref{fig:evolvedbs}, shows the separate BS and the Mobile Clone Cloud. The Base station is connected to the mobile cloud high bandwidth, low latency transport network. The mobile cloud consists of multiple Physical Machines (PMs) that host clones. The PMs are also interconnected with a high bandwidth, low latency networks. As the cloud enabled BS evolves, as shown in the right side of the Figure \ref{fig:evolvedbs}, the mobile cloud is integrated inside the mobile BS and transformed into a one entity. Such an integration removes the delay introduced by the link between the BS and the mobile cloud, and further couples computing with communications. 

\subsubsection*{Mobile Cloud Controller}

\begin{figure*}[ht]
\centering
\includegraphics[scale=0.7]{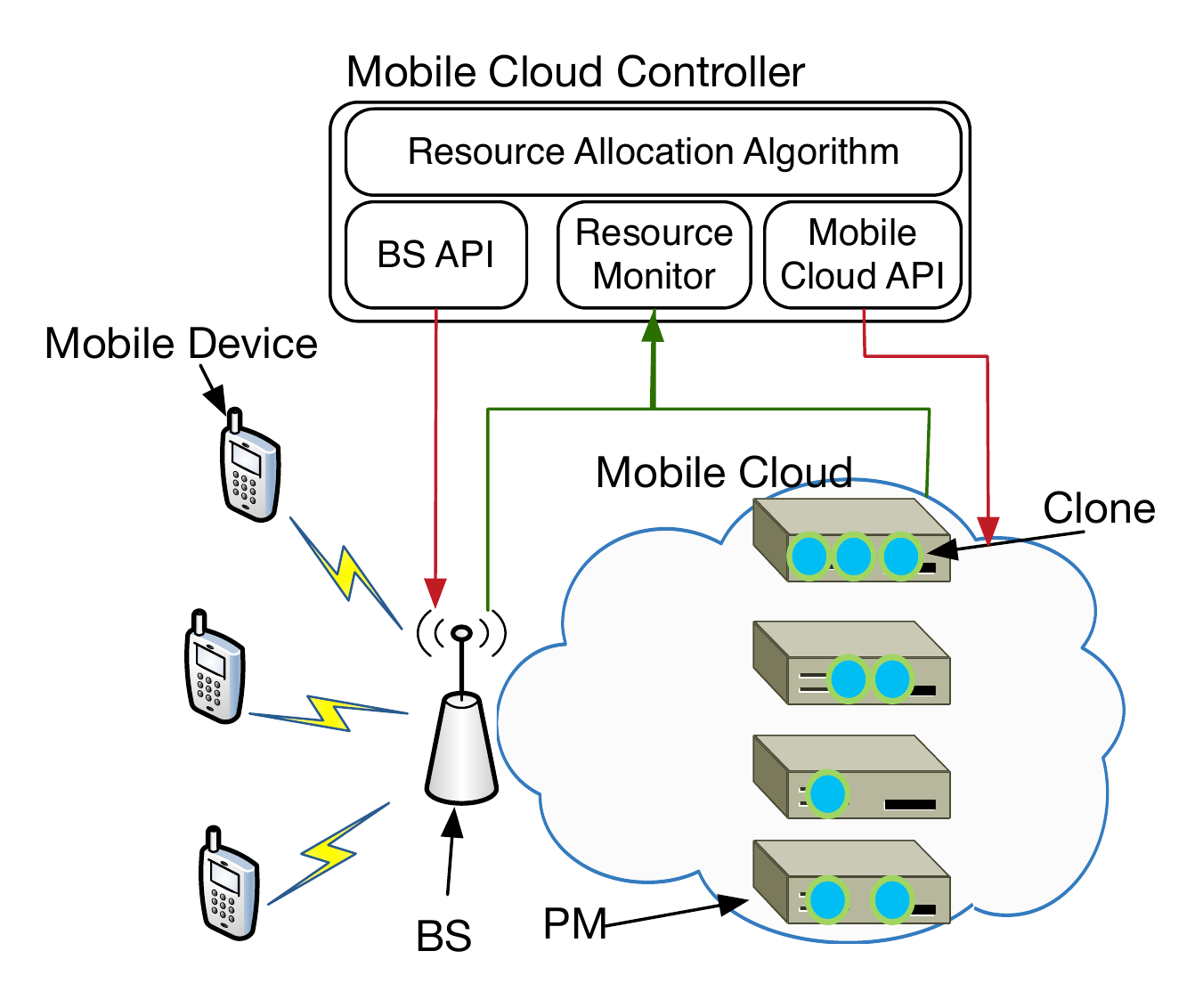}
\caption{\label{fig:mccontroller} Mobile Cloud Controller for computation dissolving}
\end{figure*}

As a result of the coupling of compute and radio resources in to the BS, a control functionality for the mobile cloud has to be introduced that is also capable of operating with existing control elements in basestations. A mobile cloud controller, in future may reside inside the evolved clone-enabled BS, although in the pre-evolved stage we assume the controller is a separate entity as depicted in Figure \ref{fig:mccontroller}. Such an approach for managing resources in multiple domains with computing resources, is called computation dissolving. The blue lines represent monitoring data flows, and the resource allocation control streams are shown in colour red.  The resource monitoring and analytical modules in the controller receives the monitoring information of both radio and clone resources. The BS API performs radio resource management tasks on the mobile network, while the module with the Mobile Cloud API performs computing resource management on the mobile clouds.

Such a resource controller will not only allow independent control of resources, also joint and cooperative resource management of multiple domains. The objective of resource management algorithms may benefit the mobile users or/and the operators simultaneously.

\section{Prototype Implementation}
\label{sec:mcprototype}

A prototype implementation has been developed by the authors for further investigating computation dissolving in mobile networks. C-RAN has been used as the mobile network deployment architecture for the prototype. The centralisation of the computational resources of C-RAN have significant benefits such as, improved load balancing, resource sharing, joint and cooperative resource scheduling, increased energy consumption by exploiting the load variations and saving the operating expenses due to centralised maintenance. 

MCC brings rich and powerful properties of cloud computing to mobile computing. Compute offloading is one of the techniques that were invented to achieve energy efficiency of mobile devices, by offloading computationally intensive tasks to a remote cloud to be executed. Existing compute offloading systems take coarse network information when making offloading decisions and the remote compute offloading target is nothing more than a computationally rich node.

Extensive research has been carried out on RAN technologies and MCC independently in literature. With the notion of Aqua computing, we believe that mobile cloud computing techniques can be well adopted to the "Clone". A mobile cloud controller has been implemented for compute resource dissolving with other resources, for joint resource allocation between mobile cloud and the BS.

\begin{figure}
\centering
  \includegraphics[scale=1]{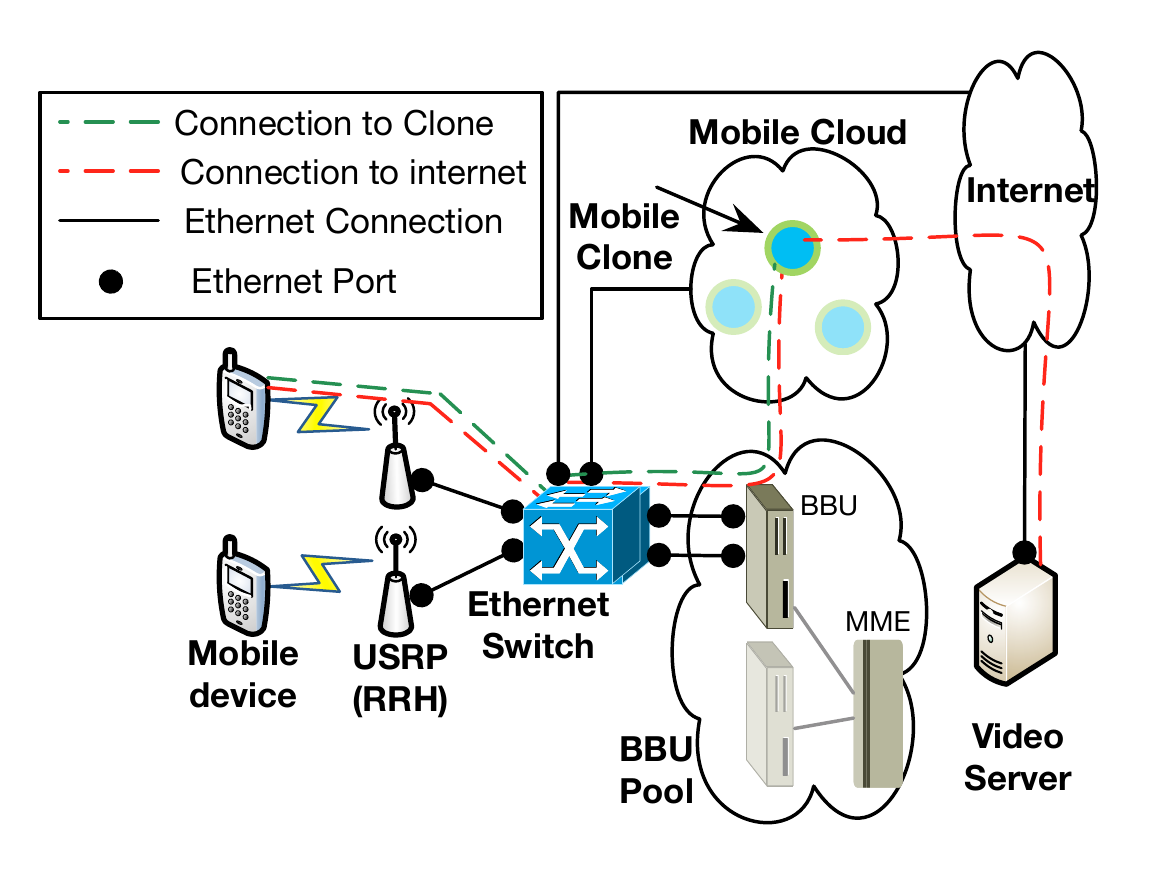}
  \caption{\label{fig:crantestbed} LTE C-RAN testbed with Mobile Cloud }
\end{figure}

The authors have been able to perform and evaluate some of use cases and applications in the developed test environment. Figure \ref{fig:crantestbed} depicts the design of the test-bed. There are USRP n210 and x300 devices devised as remote radio heads (RRH), which are connected to the BBU pool via a Gigabit Ethernet switch. Amarisoft LTE 100 and Open Air Interface (OAI) base station software are running on Dell PowerEdge R210-II rack servers. Openstack has been deployed as the cloud framework that hosts the mobile clones running Android-x86 operating system, on 13 R415 and R210-II Dell PowerEdge servers. Gigabit Ethernet links serves all network nodes. In the current setting, the wireless bandwidth has been set to 10 MHz.  The BBUs are connected to its Mobility Management Entity (MME) via its S1-MME links using S1 Application Protocol (S1AP) that provides signalling between E-UTRAN and the evolved packet core (EPC). The black solid line shows the physical Ethernet connections between nodes and networks. The green dotted line shows the logical connectivity between the mobile device and the mobile clone. The red dotted line shows the logical connectivity between the mobile device and the internet.  You can observe that via the green and red logical connections, computational and network resources are served to mobile devices collectively, for compute resource dissolving.

\subsection{Mobile Cloud Controller}

\begin{figure}
\centering
  \includegraphics[scale=1]{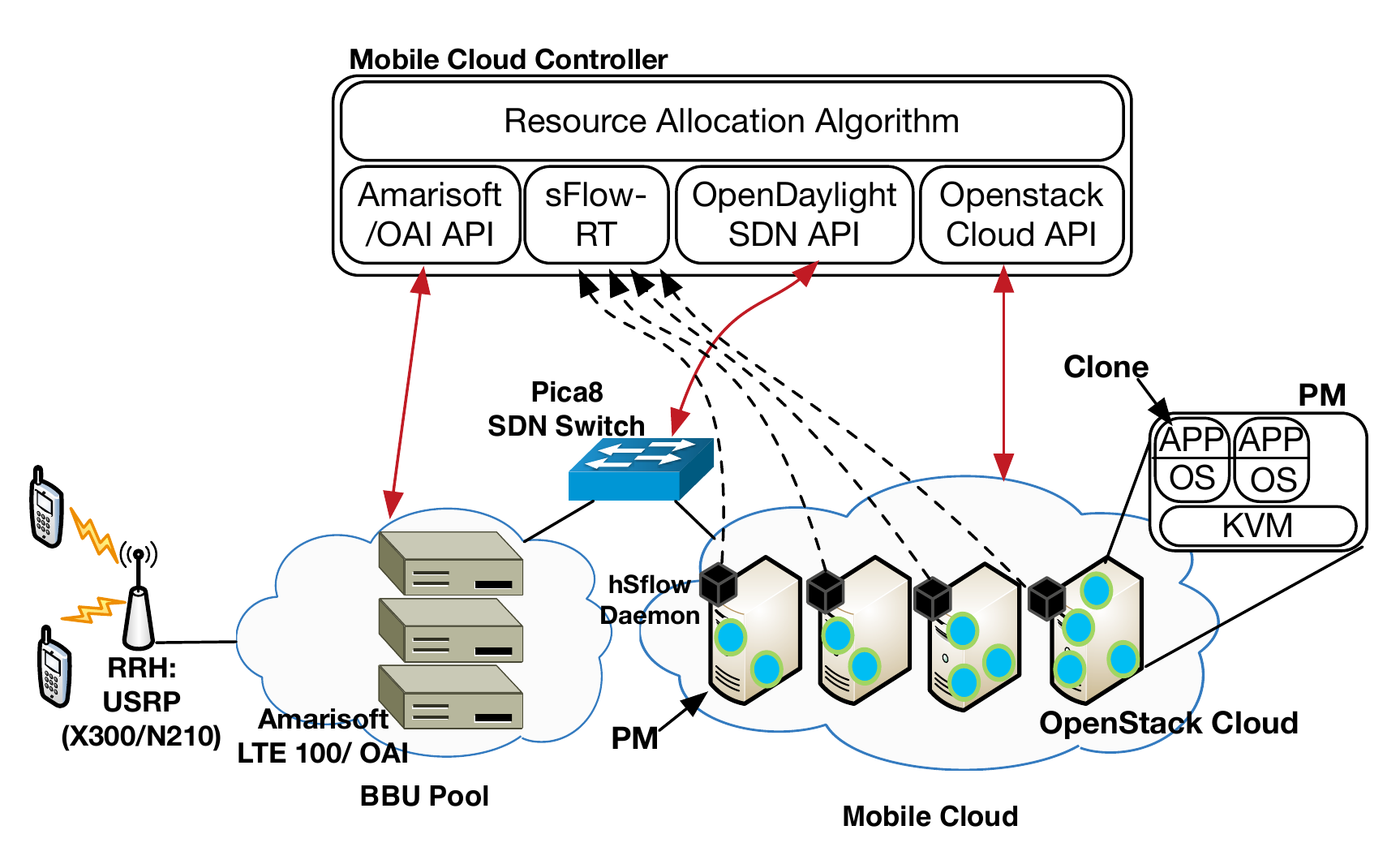}
  \caption{\label{fig:mccontrollerimpl} Mobile Cloud Controller in LTE C-RAN testbed with Mobile Cloud }
\end{figure}

One of the aspects of compute resource dissolving is joint and/or cooperative resource allocation between multiple types of resources. A separate server node is dedicated to the Mobile Cloud Controller. As shown in Figure \ref{fig:mccontrollerimpl}, the mobile controller uses multiple APIs to allocate resources in the infrastructure. The LTE APIs (Amarisoft LTE 100, OAI) are used to gather monitoring information from the RAN. Amarisoft and OAI software BSs are connected to USRP N210 and X300 USRPs over an ethernet network, which are acting as the RRHs of the network.

From the mobile cloud side, the mobile cloud controller uses sFlow protocol to receive monitoring information from the mobile cloud. Moreover, the Host-sFlow (HsFlow) daemons that are running in the mobile cloud sends periodic sampled monitoring information to Real time sFlow analytical tool in the controller. HsFlow monitors information of, physical (PM), hypervisor (KVM), virtual machine (Clone) and application (Compute offloading) levels. The openstack API is used to manage and allocate computational resources for the subscribers. A SDN switch provides connectivity between the RAN and the mobile cloud. OpenDaylight SDN controller is hosted in the mobile cloud controller, which controls the Pica8 SDN switch. The SDN switch provides loads balancing and traffic migration between the clones.

\subsection{Offloading Environment}

Openstack has been deployed as the cloud framework for orchestrating virtual resources in the cloud. In the prototype for compute offloading, a compute offloading framework called thinkair \cite{6195845} has been deployed. A modified (cloudified) version of the Android OS that runs on X86 processor architecture has been installed in the clone that the server side of Thinkair runs on. Cloudification involves adding kernel modules to the OS for utilising virtual resources that are presented by the VM, and adding modules to dynamically tailor the cone and to automate mobile cloud tasks. 

\begin{figure}
\centering
  \includegraphics[scale=1]{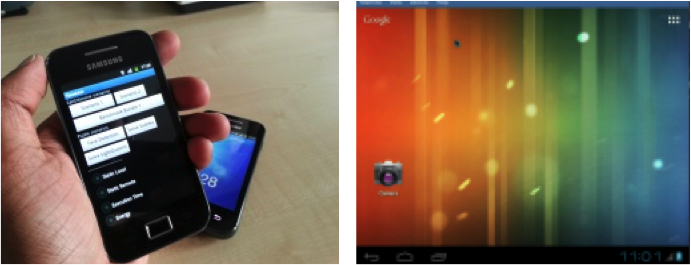}
  \caption{\label{fig:offloadingpic} Mobile compute offloading. Left: Compute offloading framework in mobile side, Right: Mobile Clone }
\end{figure}

In the mobile device's side, the mobile components of Thinkair has been installed. Running components of Thinkair, on both the clone and the mobile device as shown in Figure \ref{fig:offloadingpic}.

\subsection{RTMP streaming with Clone enabled BS}

Video streaming over RTMP has been widely used today for delivering live video streams as well as Video On Demand (VOD) services. In a conventional system, a video stream is streamed from a remote server to the client/mobile device directly. Wireless networks in comparison to wired networks are unpredictable and unstable. The stability of wireless networks depends on parameters such as the interference, the distance from the mobile device to the connected base station and congestion. Due to instabilities in wireless networks packet loss and packet retransmissions occur; in turn, it leads to increase in traffic in the backhaul networks.

\begin{figure}
\centering
  \includegraphics[]{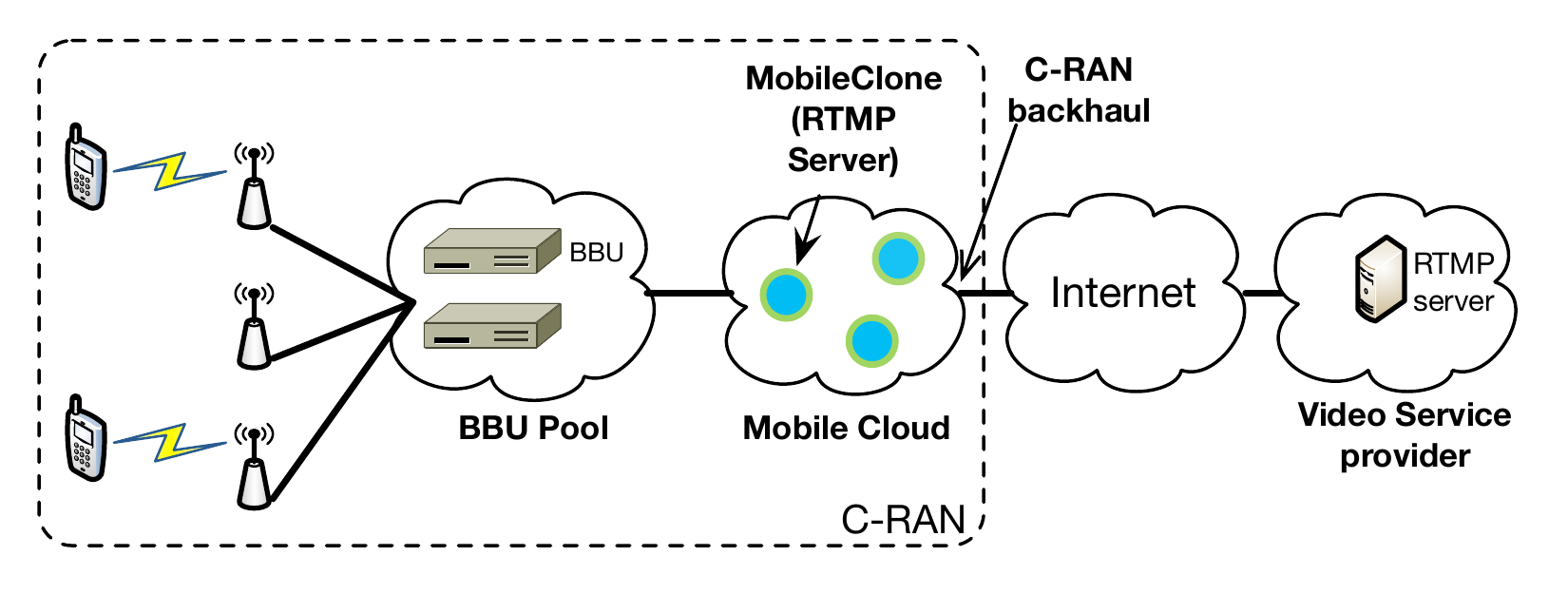}
  \caption{\label{fig:crancloudvideobuffer} C-RAN testbed architecture for video streaming with RTMP}
\end{figure}

In the proposed architecture, a mobile cloud has been placed in the mobile network as a part of the wireless network infrastructure. Similarly, to the scenarios discussed above, one clone per each mobile subscriber will be assigned. The mobile clone acts as a local buffer point such that, the mobile device gets the video stream through its clone. When issues in the wireless network such as packet loss occur, the mobile clone handles retransmissions instead of the remote video source. The buffer point that is placed at the edge of the network reduces the retransmission packet delay and reduces retransmission traffic in the backhaul network. We can conclude that, it separates wireless related issues from the backhaul networks. As shown in Figure \ref{fig:crancloudvideobuffer}, the user uses RTMP streaming for getting video content. 

As mentioned above, it is assumed that the mobile device requests a video from the video service provider through its clone. The video source responds by initiating an RTMP stream to the RTMP server running in the clone. Once the clone receives the stream from the remote video service provider, it buffers and forwards the stream back to the mobile device (possibly after some transcoding if needed).

% **********
% Sequence diagram -  video streaming.
%as illustrated in Figure 5.
% **********

When a video has been requested, the remote video server starts a video encoder, which might encode the video to the requested video quality or copy the frames as it is to the RTMP application that is setup on a Nginx streaming server. Then, it forwards the stream to the user's clone. Once the clone gets the stream, the mobile RTMP client on the mobile device can connect to it and start streaming the video.

For performing experiments, a PC with a Quad-core Intel(R) Core(TM) i5-2400 CPU @ 3.10GHz processing power and 4G RAM has been used as the remote video source and the encoder. The encoder and the video server are hosted on the same node. The clone is of type Openstack flavour tiny 1, which consists of 1 CPU and 512MB of RAM. The video that was being used for streaming is a 480p video with 1508 kb/s bitrate. To emulate packet loss that may occur due to wireless related issues such as congestion or interference in the wireless network, netem network emulation tool has been used to introduce packet loss into the wireless network in the BBU.

\begin{figure}
\centering
  \includegraphics[scale=0.44]{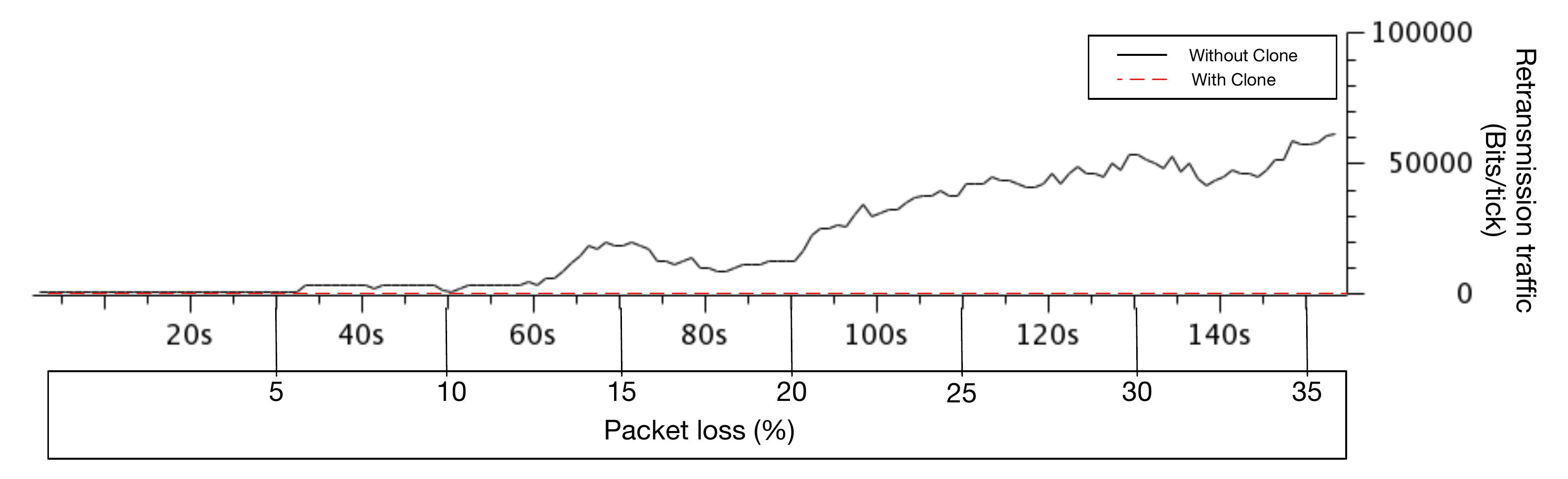}
  \caption{\label{fig:retranscomparison } Video retransmission traffic in RAN backhaul. With and without Clone}
\end{figure}

Figure \ref{fig:retranscomparison } shows how the clone has reduced RTMP retransmission traffic in the RAN backhaul network.  The straight line shows a conventional video-streaming scenario without a clone in place, we can see that as the packet loss rate has been increased, the retransmission traffic has also been increased as a result of TCP reliable stream delivery service. However, with the proposed architecture the clone that is placed next to the BBU does retransmission of RTMP packets instead of the remote video server, so there is almost no retransmission traffic in the backhaul network, as shown by the dashed line. Thence, the bandwidth of the backhaul network has been saved, as retransmission traffic does not reach the video source.

\section{Conclusion}

In this paper the authors introduce a new vision for computing, where computing resources and communication resources are coupled for enhancing future connected devices and to better optimise services. Compute-communication coupling is achieved by joint/cooperative resource allocation, hardware integration and service integration, for enhancing computing capacity of connected devices. The benefits of such a computing model are analysed by the authors from both the end users' perspective and the service providers perspective. To validate the proposed vision for aqua computing, the authors propose an architecture that reflects the characteristics of aqua Computing. In mobile cloud, at the edge of the mobile network, computation dissolving is achieved by the mobile cloud controller, while also computing and communication services are integrated by the mobile service provider. A prototype of the proposed architecture is developed, and the gathered results shows how aqua computing benefits the users and the service providers.

\ifCLASSOPTIONcompsoc
  \section*{Acknowledgments}
\else
  \section*{Acknowledgment}
\fi

This work was supported by UK EPSRC NIRVANA project (EP/L026031/1) and EU Horizon 2020 iCIRRUS project (GA-644526)

\ifCLASSOPTIONcaptionsoff
  \newpage
\fi



\bibliographystyle{ieeetr}


%\vfill


\end{document}